\documentclass[%
 aip,
 amsmath,amssymb,
 reprint,%
]{revtex4-1}

\usepackage{graphicx}% Include figure files
\usepackage{dcolumn}% Align table columns on decimal point
\usepackage{bm}% bold math
\usepackage{placeins} 
\usepackage[utf8]{inputenc}
\usepackage[T1]{fontenc}
\usepackage{inconsolata}
\usepackage{comment}

\usepackage{xcolor}
\definecolor{myblue}{RGB}{0, 206, 209}
\definecolor{mypink}{RGB}{255, 0, 255}
\definecolor{myorange}{RGB}{255, 172, 28}

\usepackage{etoolbox}
\usepackage{booktabs}
\usepackage{textcomp}
\usepackage{comment}
\usepackage{hyperref}
\makeatletter
\def\@email#1#2{%
 \endgroup
 \patchcmd{\titleblock@produce}
  {\frontmatter@RRAPformat}
  {\frontmatter@RRAPformat{\produce@RRAP{*#1\href{mailto:#2}{#2}}}\frontmatter@RRAPformat}
  {}{}
}%
\makeatother
\begin{document}

%\preprint{AIP/123-QED}

\title[]{How we simulate DNA origami}

\author{Sarah Haggenmueller}
\affiliation{
School of Natural Sciences, Department of Bioscience, Technical University Munich, 85748 Garching, Germany.
}%

\author{Michael Matthies}
\affiliation{
School of Natural Sciences, Department of Bioscience, Technical University Munich, 85748 Garching, Germany.
}%

\author{Matthew Sample}

\affiliation{School for Engineering of Matter, Transport, and Energy, Arizona State University, Tempe, AZ 85287, USA}

\affiliation{ 
School of Molecular Sciences and Center for Molecular Design and Biomimetics,
The Biodesign Institute, Arizona State University,
1001 South McAllister Avenue, Tempe, AZ 85281, USA
}%
\author{Petr \v{S}ulc}%

\affiliation{
School of Natural Sciences, Department of Bioscience, Technical University Munich, 85748 Garching, Germany.
}%
\affiliation{ 
School of Molecular Sciences and Center for Molecular Design and Biomimetics,
The Biodesign Institute, Arizona State University,
1001 South McAllister Avenue, Tempe, AZ 85281, USA
}%

\affiliation{Center for Biological Physics, Arizona State University, Tempe, AZ, USA}

\date{\today}% It is always \today, today,

\begin{abstract}
DNA origami consists of a long scaffold strand and short staple strands that self-assemble into a target 2D or 3D shape. It is a widely used construct in nucleic acid nanotechnology, offering a cost-effective way to design and create diverse nanoscale shapes. With promising applications in areas such as nanofabrication, diagnostics, and therapeutics, DNA origami has become a key tool in the bionanotechnology field. Simulations of these structures can offer insight into their shape and function, thus speeding up and simplifying the design process. However, simulating these structures, often comprising thousands of base pairs, poses challenges due to their large size. OxDNA, a coarse-grained model specifically designed for DNA nanotechnology, offers powerful simulation capabilities. Its associated ecosystem of visualization and analysis tools can complement experimental work with in silico characterization. This tutorial provides a general approach to simulating DNA origami structures using the oxDNA ecosystem, tailored for experimentalists looking to integrate computational analysis into their design workflow.
%computational design and simulation methods 
\end{abstract}

\maketitle

\section{Introduction} 
DNA origami is one of the most widely used constructs in nucleic acid nanotechnology \cite{dey2021dna}. Since its introduction by Paul Rothemund \cite{rothemund2006folding} for 2D shapes, the techniques have been adapted to a variety of more complex shapes in 3D \cite{douglas2009self}. The versatility of DNA, which includes functionalization with chemical groups such as silicification or metallization \cite{madsen2019chemistries} and the ability to cross-link with cargo like gold nanoparticles, proteins, and small molecules, makes DNA origami a highly flexible platform. It allows for nanoscale positioning, with applications extending into plasmonics, photonics, diagnostics, and therapeutics \cite{}. 
With the growing popularity of DNA origami, a range of design and computational modeling tools have been developed. One of the first tools, which still remains highly popular, is the caDNAno software \cite{douglas2009rapid}, which uses a 2D mesh to aid in the design and placement of staple strands. A new generation of tools, such as Athena \cite{jun2021rapid}, Adenita \cite{de2020adenita}, MagicDNA \cite{huang2021integrated}, ENSNano \cite{levy2021ensnano}, and DNAForge \cite{elonen2024dnaforge}, started to emerge in the past few years. Those allow users to specify the shape of the target structure and aim to (semi)automate the design of DNA origami. With the growing use of DNA origami and the development of more complex designs, the need for reliable simulations of DNA nanostructures has also grown. These simulations help to assess if the nanostructures fold into the desired shape and to evaluate their flexibility.
Existing tools include CanDO \cite{kim2012quantitative}, SNUPI \cite{lee2022predicting} and MrDNA \cite{maffeo2020mrdna}. We focus here on the oxDNA package \cite{snodin2015introducing,doye2013coarse,poppleton2023oxdna}, co-developed by members of our team, as it is the only model capable of explicitly handling the breaking and forming of DNA duplexes. This makes it ideal for simulating complex designs with both single- and double-stranded regions, as well as capturing phenomena like entropy-driven bending of DNA origami tiles \cite{sample2023hairygami,yu2023cytodirect}. The oxDNA ecosystem includes several complementary tools, such as the oxView visualizer \cite{bohlin2022design,poppleton2020design}, the free simulation webserver oxDNA.org \cite{poppleton2021oxdna}, and a Python library for analyzing simulation results \cite{poppleton2023oxdna}. Additionally, it supports an associated online structure database \cite{poppleton2022nanobase} enabling researchers to share and collaborate on designs. As opposed to fully atomistic simulations, the coarse-grained oxDNA model is fast enough to capture the dynamics of DNA origami structures. However, it is not efficient enough to simulate the entire assembly process from single strands without relying on time-consuming advanced simulation techniques. For such large-scale assembly tasks, even coarser models are required \cite{deluca2024mechanism,dannenberg2015modelling}.

In this tutorial we will give an introduction on how to simulate simple DNA origami structures using the oxDNA ecosystem, starting from structure import to setting up a simulation, as well as analyzing it. In order to do so, we have to understand some key concepts of the structural design and the algorithms used in the model:

\textbf{DNA origami:} DNA origami is the self-assembly of a long single-stranded DNA molecule called 'scaffold' into specific shapes using short complementary strands called 'staples'. These staples bind the scaffold at precise locations, guiding it to fold into a desired structure \cite{rothemund2006folding, douglas2009self}.

\begin{figure}[t]
    \centering
    \includegraphics[width=\linewidth]{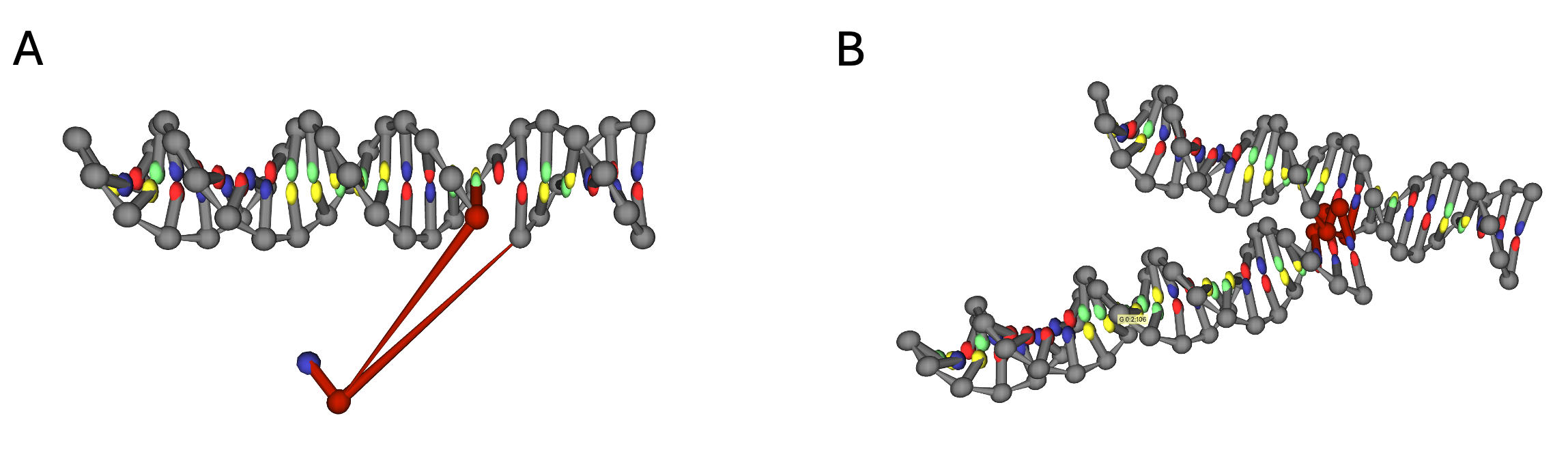}
    \caption{\textbf{Overstretched bonds and overlapping nucleotides.} (A) Highlighted in red are overstretched bonds that can result from a misplaced nucleotide. (B) Highlighted in red are overlapping nucleotides that might cause problems during relaxation.}
    \label{fig:overlap_overstretch}
\end{figure}

\textbf{Overstretched bonds and overlapping:} Due to improper initial geometries that can result from conversion of the design files, some molecules can be positioned too close (overlapping) in respect to the neighbouring nucleotides or too far (overstretched bonds) in respect to other nucleotides (Figure \ref{fig:overlap_overstretch}). Both phenomena are non-physical configurations and lead to instability of the molecular dynamics simulations of the structure.

\textbf{Relaxation and Equilibration:} Before the behavior of a structure can be simulated, it has to have reached a stable state, where no highly unfavorable interactions such as overlapping particles are present. Through relaxation and equilibration any unrealistic forces or distortions in the initial structure are removed until the system reaches the desired stable state\cite{bohlin2022design}.

\textbf{Monte Carlo simulation:} Monte Carlo (MC) is an iterative process that samples a system's configuration by generating new states through random movements or rotations of nucleotides. These changes are then accepted or rejected based on specific criteria that depend on the overall change of energy that such a move would generate. When running MC at low temperatures, the algorithm is very likely to accept moves that lower the overall energy, allowing the system to progressively relax towards a lower-energy, stable configuration \cite{torrie1974monte, Metropolis1949montecarlo}. This method is particularly useful during early stages of relaxation, helping to resolve issues like massively overstretched bonds and nucleotide overlaps, which are unfavorable high-energy states.

\textbf{Molecular Dynamics simulation:} Molecular Dynamics (MD) allows the positions of atoms and molecules to evolve under forces derived from potential energy functions \cite{Yoo2013MD}. Through these motions, overstretched bonds and distorted atomic arrangements are corrected, gradually leading towards a stable equilibrium state\cite{Doye2023}. However, extremely overstretched bonds generate high forces that can cause atoms to move too quickly. This can cause numerical instabilities, potentially causing the simulation to crash. As opposed to MC, MD can be easily parallelized on GPU, and is thus much faster to run, but its numerical integrator is prone to instabilities if very high energy states (like overstretched bonds) are present.

\section{Results and Discussion}
\begin{figure*}[t]
    \centering
    \includegraphics[width=\linewidth]{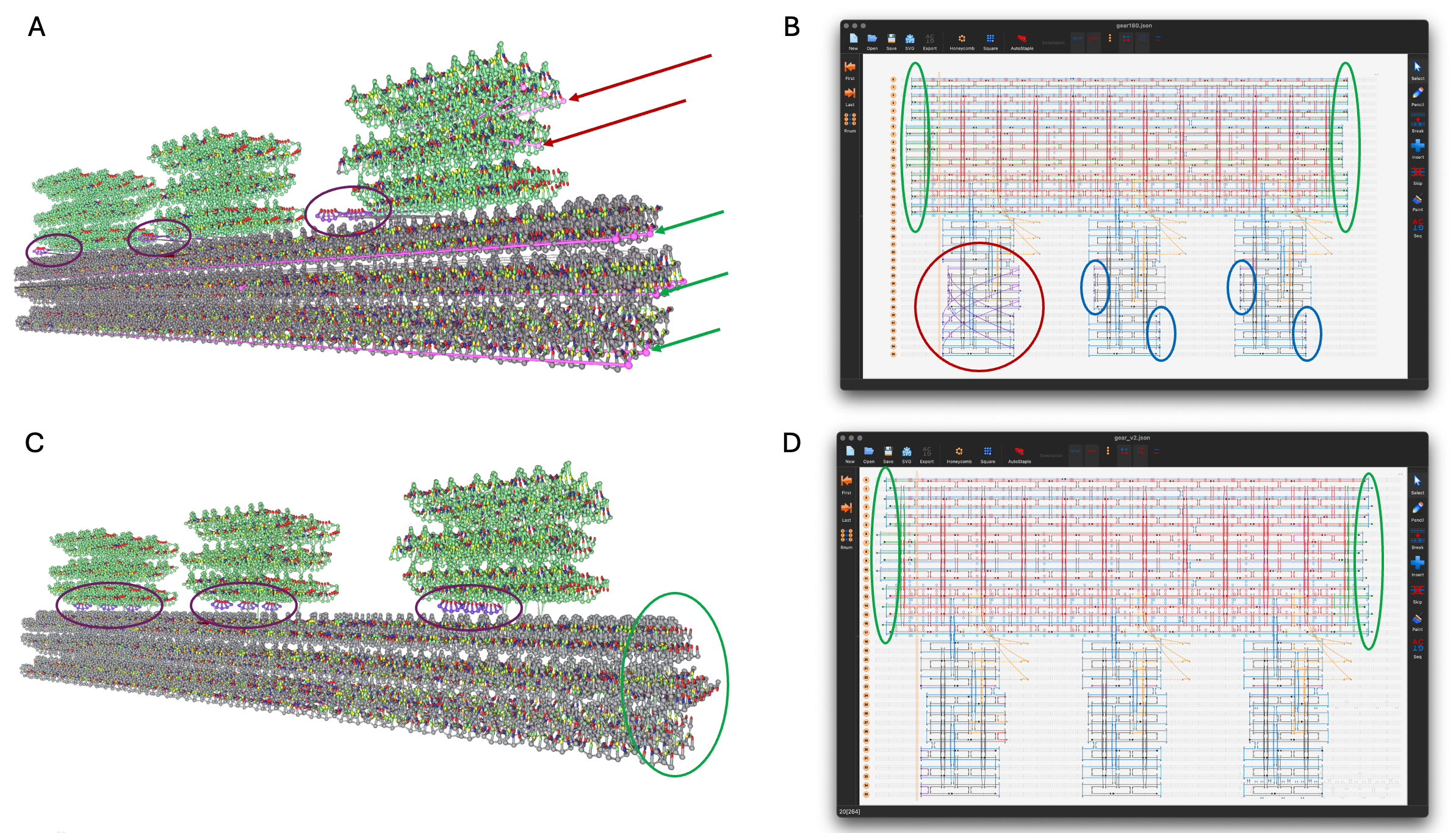}
    \caption{\textbf{Modifications in caDNAno that allow for proper relaxation of the gear180.} (A) Original structure visualized in oxView. Coloring was changed ('View' > 'Colors') for better visualization of the base part (grey), the 3 bricks (green) and the appendices of the bricks (purple). Overstretched bonds (pink) that are spanning the base (green arrows) and one brick (red arrows) have to be modified before relaxation. (B) Screenshot of caDNAno that visualizes problematic areas. Highlighted are overstretched bonds that span the structure (green) and one of the bricks (red). Highlighted in blue are very short staple sequences that would easily detach and that should be removed as well. (C) Modified structure visualized in oxView. Overstretched bonds were removed and the three appendices (purple) were moved between bricks and base. Highlighted in green are the staple sequences that were re-added and assigned as thymines when loading into oxView. (D) caDNAno view of the structure after modifications were made. Highlighted in green are the areas of cut and re-added staple sequences.}
    \label{fig:modifications}
\end{figure*}

\subsection{Before we simulate}
Before starting a simulation, it is important to consider what kind of information we aim to extract from the simulations and ensure that the chosen model is accurate enough to provide meaningful insights into the structure’s behavior. We focus here on the oxDNA model, which has been shown to quantitatively reproduce the behavior of both single- and double-stranded DNA systems 
\cite{ouldridge2011structural,snodin2015introducing} as well as DNA origami structures \cite{snodin2019coarse,doye2013coarse,Doye2023}. Even though the model only uses a Debye-H\"{u}ckel approximation parameterized to DNA thermodynamic data across different sodium concentrations \cite{snodin2015introducing}, it has been shown to be in agreement with observed structures under high salt conditions, both in monovalent and divalent buffers \cite{sample2023hairygami}. It cant represent DNA, RNA, hybrid DNA-RNA systems\cite{ratajczyk2023dnarnahybrid}, as well as protein-DNA or protein-RNA systems \cite{Bohlin2022dnaprotein,procyk2021coarse} with simple representation of proteins just as rigid bodies with excluded volume. Many experimental systems cannot be fully represented by this model, such as complex interactions with charged surfaces, lipids, protein or small molecule binding, electrostatic interactions with gold nanoparticles, or chemical modifications like metallization or silicification.
Additionally, oxDNA does currently not support non-canonical base pairs, such as G-quadruplexes or Hoogsteen pairing, making it unsuitable for simulating structures like DNA triplexes. Hence, if your system includes these complex motifs or interactions, oxDNA simulations will not fully capture them.

On the other hand, successful examples where oxDNA can significantly assist experiments or provide valuable design feedback include studies of systems under force, tension, or twist and their mechanical effects on DNA or nanostructures \cite{ouldridge2011structural,engel2018force,wong2022characterizing}, dynamic processes like strand displacement \cite{srinivas2013biophysics}, flexible systems capable of adopting multiple conformations \cite{wong2022characterizing}, as well as complex multi-DNA origami interacting systems \cite{yao2020meta,liu2024inverse}.  The complexity of the system or reaction directly impacts the setup and analysis of the simulations. Dynamic systems that undergo significant reconfiguration, such as strand displacement involving the breaking and forming of hydrogen bonds, often require advanced biasing techniques like forward flux sampling or umbrella sampling. While these methods are beyond the scope of this article, they are discussed in more detail in Refs.~\cite{sample2023hairygami,sengar2021primer} and are well-documented in the oxDNA package documentation \cite{poppleton2023oxdna}. 

The following section outlines the procedure for setting up and running oxDNA simulations of two different DNA origami structures without biasing. These simulations can still provide helpful insight on structure shape and conformations and help determine if the equilibrium state matches the intended design. We begin with a simple brick structure and progress to a more complex bend structure, addressing potential issues along the way. The relaxation and equilibration of structures will be conducted using the web-based oxView.org software. Subsequently, the structures will be submitted for production runs on oxDNA.org, and the results will be analyzed to ensure the simulations are sufficiently long and accurately represent the system's experimental behavior. The procedure is designed to be user-friendly and accessible, requiring no software installation. Following this tutorial, experimentalists should be able to easily perform the simulations without additional setup.

%Indeed, it is important to first outline what one wants to learn from the simulation and if the simulated system is accurately captured by the used model. 

\subsection{Structure import and modifications}

% Details on how to use this service can be found in (Poppleton et al. 2021)\cite{poppleton2021oxdna}.

The first step for setting up simulations of DNA origami is a nanostructure design, for which users typically use several of the currently available tools. To-date, the arguably most popular tool remains caDNAno (or a similar web-based tool sCADNano \cite{doty2020scadnano}). For caDNAno and Tiamat \cite{williams2009tiamat}, one needs to use the TacoxDNA tool \cite{suma2019tacoxdna} to convert the file format into the oxDNA file format (.top and .dat files). Many other tools for DNA nanostructure design (such as ENSnano,  MagicDNA, DNAforge or Adenita) support direct export into oxDNA format. Other tools that support export to PDB (such as Athena) can also be imported into the oxDNA file format through TacoxDNA. Since the majority of labs still rely on caDNAno designs, we explicitly review here the import procedure from this file format. The relaxation steps following the structure import are however similar for all the design tools.

Starting with design files from caDNAno, we firstly visualize the structures in oxView. 
In our example, we start with the 'i\_16x4.json' file, which can be downloaded from \href{https://cadnano.org/index.html}{cadnano.org} under 'gallery' > 'download source files: zip' of 'Rapid prototyping of 3D DNA origami shapes with cadnano'\cite{douglas2009rapid} and then downloading the file from the dropbox. 
To import the JSON file, we will use the 'TacoxDNA importer', which converts the '.json' file into a format compatible with oxView. 
In \href{https://sulcgroup.github.io/oxdna-viewer/}{oxview.org}, under 'File' > 'Import' > 'Taco', select the '.json' file, set the file format to 'caDNAno', choose the lattice type and 'set sequence' depending on which scaffold was used for the design. In our example the structure was designed with a 'hexagonal' lattice and the 'p8064' scaffold. If unsure, the file can be opened in \href{https://scadnano.org}{https://scadnano.org}, where the visualization on the left is indicating a hexagonal or squared lattice design.

\begin{comment}
\begin{figure}[t]
    \centering
    \includegraphics[width=\linewidth]{img/original_c_oxView.png}
    \caption{\textbf{Original structure visualized in oxView.} Coloring was changed ('View' > 'Colors') for better visualization of the base part (grey), the 3 bricks (green) and the appendices of the bricks (purple). Bonds spanning the base and one of the bricks (pink) have to be removed before relaxation steps.}
    \label{fig:OriginalStructure}
\end{figure}

\begin{figure}[t]
    \centering
    \includegraphics[width=\linewidth]{img/modified_c_oxView.png}
    \caption{\textbf{Modified structure visualized in oxView.} Bonds crossing the base (grey) and the right brick (green) were removed. The three appendices (purple) were moved between bricks and base. Structure is now ready for relaxation steps.}
    \label{fig:ModifiedStructure}
\end{figure}
\end{comment}

We do the same now with our second file 'gear180.json' from \href{https://cadnano.org/index.html}{cadnano.org} > 'gallery' > 'download source files: zip' of 'Folding DNA into Twisted and Curved Nanoscale Shapes' \cite{Dietz2009twistedandcurved} but select the 'M13mp18' scaffold.
When examining the structure we see one long base part that will later reshape into a half-circle and 3 bricks on top of that (Figure \ref{fig:modifications}A). 
Special attention should be paid to the overstretched bonds crossing the structure. 
These bonds are part of the staple strands designed to dimerize the structure by binding the sides of two different monomers. 
However, during monomer relaxation they are problematic, as they create extreme forces that hinder proper relaxation and equilibration. 
Therefore, these staples must be partially removed and re-added later during the final dimerization process or nicked. 
Additionally, some bonds spanning one of the bricks should be removed, which would also interfere with subsequent steps. 
While these modifications can be made in oxDNA, it is easier to perform them in caDNAno. 
In caDNAno, staples spanning the structure are indicated by green dotted lines (Figure \ref{fig:modifications}B). 
We remove the shorter ends of these staples and add the corresponding number of nucleotides to the other end. 
It is important to note that the re-added bases do not have an assigned nucleotide sequence. 
When the structure is later loaded into oxView, these bases will automatically be designated as thymine ('T'). 
This automatic assignment is acceptable for our monomer relaxation process. 
However, it becomes problematic when attempting to dimerize the structure.%, and the sequence needs to be manually assigned in oxView.
%For an in-depth tutorial on how to use caDNAno visit \url{https://cadnano.org/docs.html}. 
Alternatively, we can also cut the overstretched bonds in oxView. By selecting the base located 5' of a bond we can 'Nick' the bond in the 'Edit' tab. 
Doing so there are very short staple sequences remaining that might cause problems as they detach easily. Those 1-3 nt long sequences should in this case also be removed. 
Once all the troublesome staples are modified, we can make some final adjustments in oxView to expedite the relaxation process. 
The bricks are connected to the base part of the structure through single-stranded extensions. Those are currently misaligned and should be positioned between the base and bricks. 
To correct this, we can use the box-select tool to highlight the bricks and move them upward ('Edit' > 'Translate'). 
Subsequently, the appendices are positioned into their correct location, and the bricks lowered back into place. 
Do not position the elements too close to avoid overlaps. 
With the removed bonds spanning the base and one of the bricks, and with the re-positioned appendices the structure is now ready for relaxation (Figure \ref{fig:modifications}C-D). 
Make sure to save the progess regulary by clicking on 'File' > 'Save' > 'oxView'. 
The downloaded file can be directly drag-and-dropped into oxView or opened again by 'File' > 'Open'.

%Before being able to simulate a structure, it might need some modifications. For example, structures designed in caDNAno may include staples that are unlikely to bind effectively or are not suitable for accurate simulation.  It is important to modify these structures before proceeding with relaxation, equilibration, and simulation steps. Such modifications not only ensure proper modeling but can also significantly accelerate the relaxation step, saving time and resources.

\subsection{Relaxation Steps}
\begin{figure}[t]
    \centering
    \includegraphics[width=\linewidth]{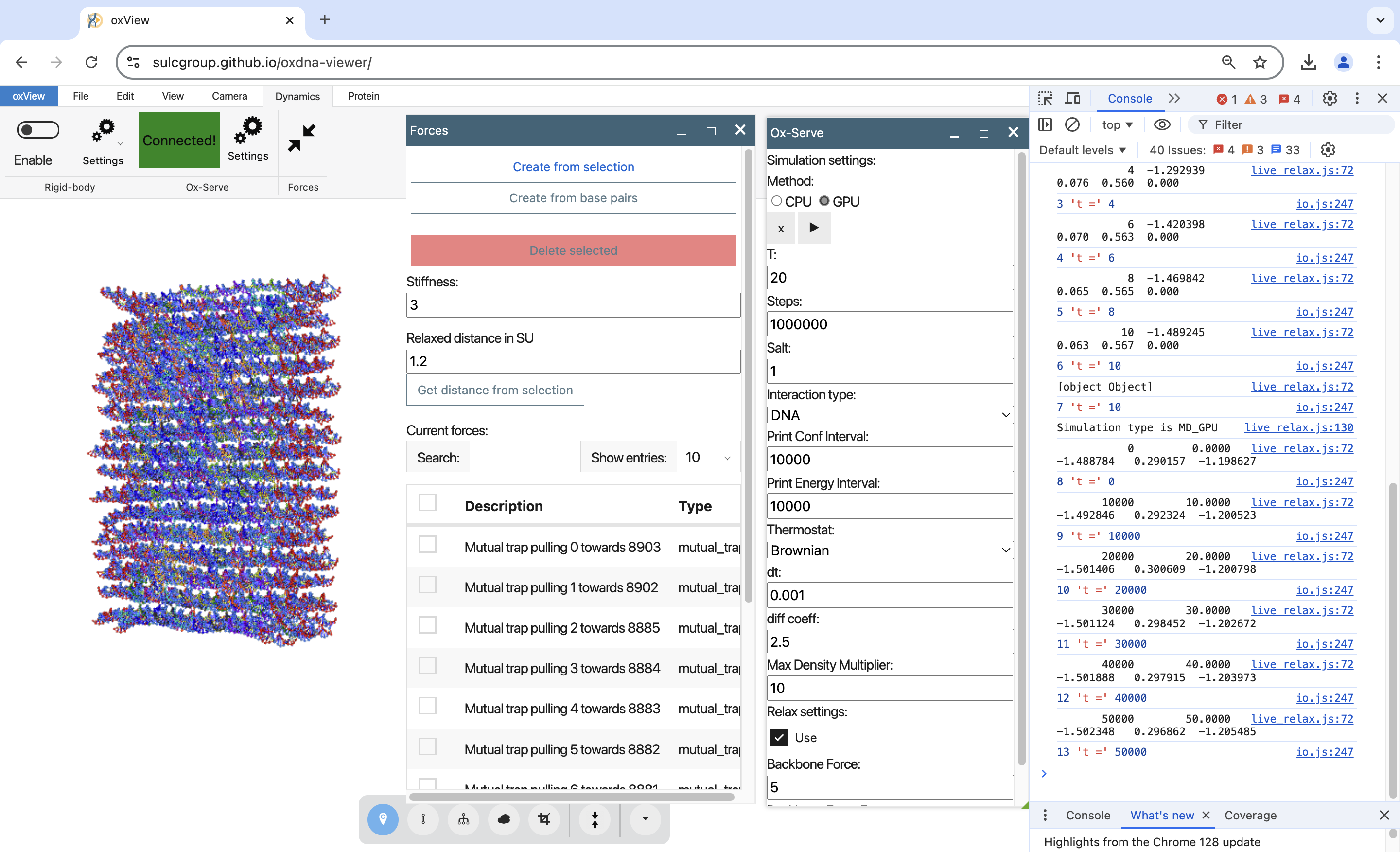}
    \caption{\textbf{Screenshot of oxView while relaxing the brick structure.} The brick structure is visualized live (left). Mutual trap forces with stiffness = 3 were applied between the base pairs and the MD relaxation protocol is run (middle) while the outputs are monitored in the console (right).}
    \label{fig:brick_relaxation}
\end{figure}

Prior to the production run, it is essential to relax the structure, as nucleotides may occupy positions that are physically unrealistic. 
Stretched bonds would generate nearly infinite forces, while overlapping nucleotides would result in significant repulsive forces in oxDNA. 
To address these issues, we begin with a few steps of Monte Carlo (MC) simulation on the CPU to resolve volume clashes caused by overlapping nucleotides. 
To improve stability of the structure it is critical to add some mutual trap forces between the base pairs during relaxation. 
Those external forces, which are implemented as a harmonic potential between interacting base pairs, keep paired bases together and help to withstand the high forces that are present in the initial structure and that would otherwise rip these base pairs apart. 
Under the 'Dynamics' tab, navigate to 'Forces', where a small pop-up window will appear. Set the 'Stiffness' value to 3.0 and click on 'Create from base pairs.' This will automatically populate the list of forces with mutual trap forces between the base pairs.
With these forces established, we can now relax the structure by running a simulation. Staying within the 'Dynamics' tab, click on 'Connect' and then on the small arrow to the left of 'NANOBASE.org' to connect via web sockets with our public oxServe host, where we offer the option to freely run the simulation on our remote server. Alternatively, if oxDNA was installed locally you can also connect to your local host following the instructions on \href{https://github.com/sulcgroup/ox-serve}{https://github.com/sulcgroup/ox-serve}. A green highlighted box in the top bar will confirm a successful connection to the server.
Next, click on 'Settings' to open another pop-up window containing the simulation settings. Here, we stay with the 'CPU' as method, indicating a MC relaxation. We set the temperature 'T' to 20°C. You can also define the number of steps the simulation should run, as well as specify the frequency for outputting energy data ('Print Energy Interval') and the frequency of the configuration update ('Print Conf Interval'). While it is nice to get feedback about the energies and current configurations frequently, it will slow down the relaxation. Therefore, ideally a number is chosen that gives you just enough information for assessing the success of the relaxation.
For our brick structure, 10 steps, with energy and configuration files printed every second step, are sufficient. Structures with multiple overstretched bonds or potential overlaps will need more MC relaxation steps. 
With 'Delta Translation' and 'Delta Rotation' we can adjust the maximal distances of each move per particle. While too small values here will slow down the relaxation process, too large values might lead to a lower acceptance of the proposed moves and thus also slow down the relaxation. It is recommended to keep those values at their default setting '0.22', as they typically lead to a reasonable move acceptance ratio. Choosing much smaller values will result in a very slow relaxation, and choosing too large values will lead in most MC moves being rejected.
Finally, make sure to tick the checkbox under 'Relax settings'. This option is crucial as it ensures that interaction potentials for backbone interactions are limited (i.e. cannot reach infinity) despite the presence of stretched bonds in the initial structure.
At this stage, it is recommended to open the developer console in your browser (key F12 for most browsers). 

\begin{figure}[t]
    \centering
    \includegraphics[width=\linewidth]{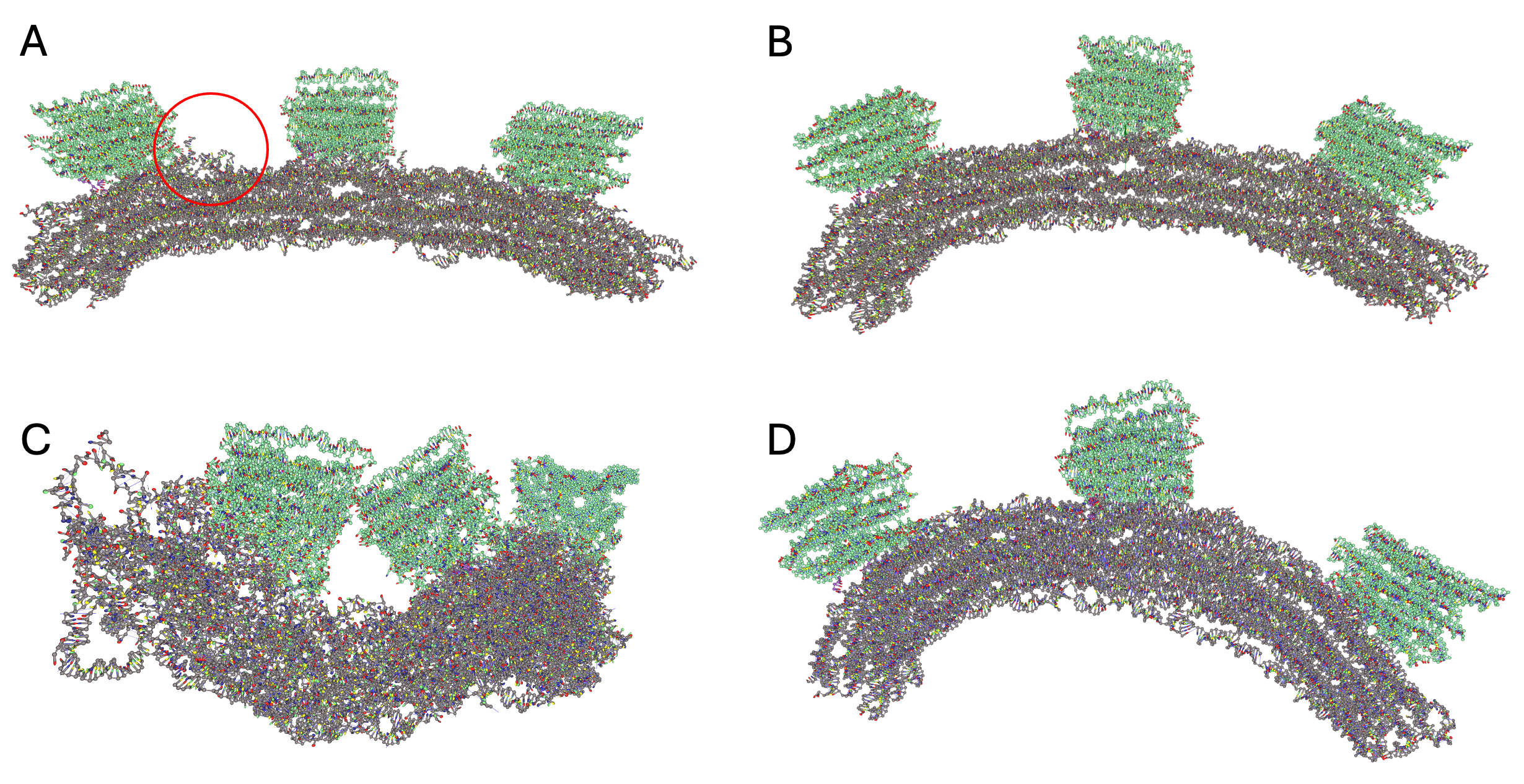}
    \caption{\textbf{Comparison of structures with different relaxation parameters.} (A) The structure was relaxed using 10 MC steps, followed by 1e6 MD steps without any additional mutual trap forces. The red circle highlights an area where parts of the structure are breaking off. (B) Increasing the MC steps to 100 increases stability and prevents the breaking off of parts during MD even without mutual trap forces in place. (C) Structure was relaxed with 100 MC steps, followed by 1e6 MD steps. Here the recommended stiffness = 3 for mutual trap forces between base pairs is preventing the structure to relax. (D) Reducing the stiffness of the trap forces to 1 enhances the relaxation process and speeds it up compared to using no trap forces, as seen by a greater bend after the same number of relaxation steps.}
    \label{fig:gear180_relaxParams}
\end{figure}

Next, click on the arrow under 'Method' to initiate the relaxation process. As the relaxation progresses, the developer console will display outputs, including the following metrics: [time (steps)], [potential energy], [acceptance ratio for translational moves], [acceptance ratio for rotational moves], and [acceptance ratio for volume moves]. Volume moves are never used in the relaxation process, so the acceptance ratio is 0. 
% For optimal results, \textcolor{red}{the acceptance ratios should be around 30\%,} and 
The potential energy should decrease over time, indicating the structure to become more stable. Reaching a value of about -1.4, we can transition to the more efficient Molecular Dynamics (MD) relaxation on the GPU (Figure \ref{fig:brick_relaxation}). Keeping the temperature 'T' set at 20°C, we adjust the integration time step 'dt' to 0.001. This smaller time step helps reduce numerical instabilities caused by the initially high forces. As before, we can specify the number of steps and print intervals, ensuring that the 'Relax settings' checkbox remains selected.
The number of steps required to relax a structure depends heavily on its design. Structures with twists generally require more steps, while smaller, linear structures relax more quickly. For this example, we’ll use 1e8 steps for relaxation. A good rule of thumb is to set 'Print Conf Interval' to steps/50 which would be 2e6 in this case and 'Print Energy Interval' to steps/100 (1e6 here). This setup will generate 100 energy outputs during the MD relaxation process. The energy output in MD includes the following metrics: [time (steps * dt)], [potential energy], [kinetic energy], and [total energy]. When the energy values do not change much anymore and the structure also visually does not seem to move much anymore we can proceed to the next relaxation step. During relaxation it is crucial to inspect the structure and ensure that no parts are breaking off, as these fragments could interfere with the simulation later on. If breakage occurs, you can address it by adjusting the mutual trap forces, lowering the maximum backbone forces, or decreasing the time step dt.

\begin{figure}[t]
    \centering
    \includegraphics[width=\linewidth]{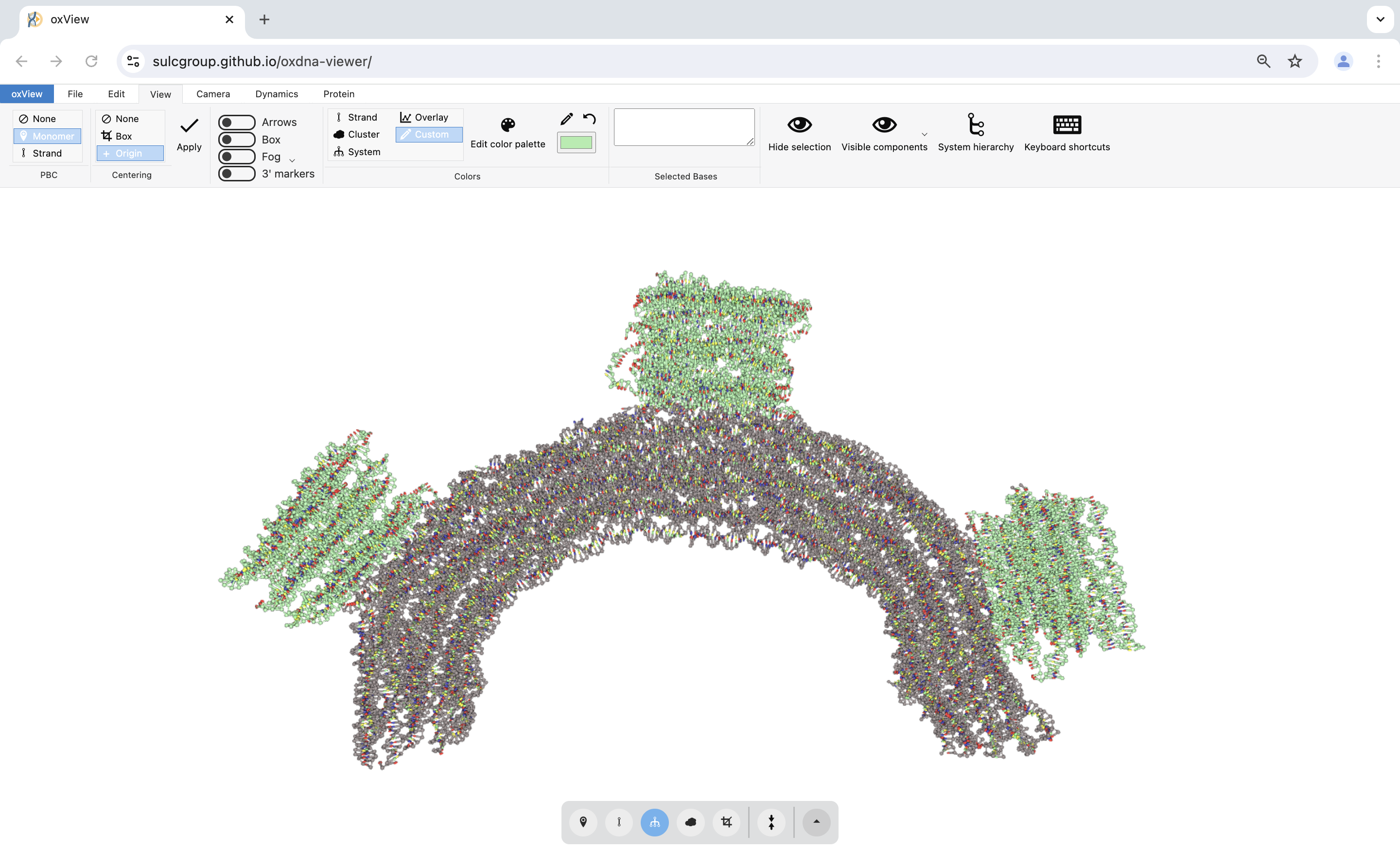}
    \caption{\textbf{Relaxed structure of gear180 visualized in oxView.} Base (grey) is now bent into a half-circle with the three bricks (green) sitting on top.}
    \label{fig:gear180relaxed}
\end{figure}

When we try the same parameters with our gear180 structure, we are encountering difficulties with the relaxation. Instead of relaxing into a half-circle, the structure is deforming (Figure \ref{fig:gear180_relaxParams}C).
Through trial and error it was determined that for this structure, it is better to relax the structure with a lower stiffness or even without having any mutual trap forces in place. 
Additionally, as this structure is more complex and there were more overstretched bonds, the MC relaxation steps have to be increased to 100 before switching to MD.
Following the protocol with those modifications, the structure should relax into a half-circle (Figure \ref{fig:gear180relaxed}). Without mutual trap forces we have to be particularly careful of broken-off nucleotides. Visual inspection of the structure is crucial here. 
If the structure appears to be stable and intact, we can proceed to the next steps.

\begin{figure}
	\centering
	\includegraphics[width=\linewidth]{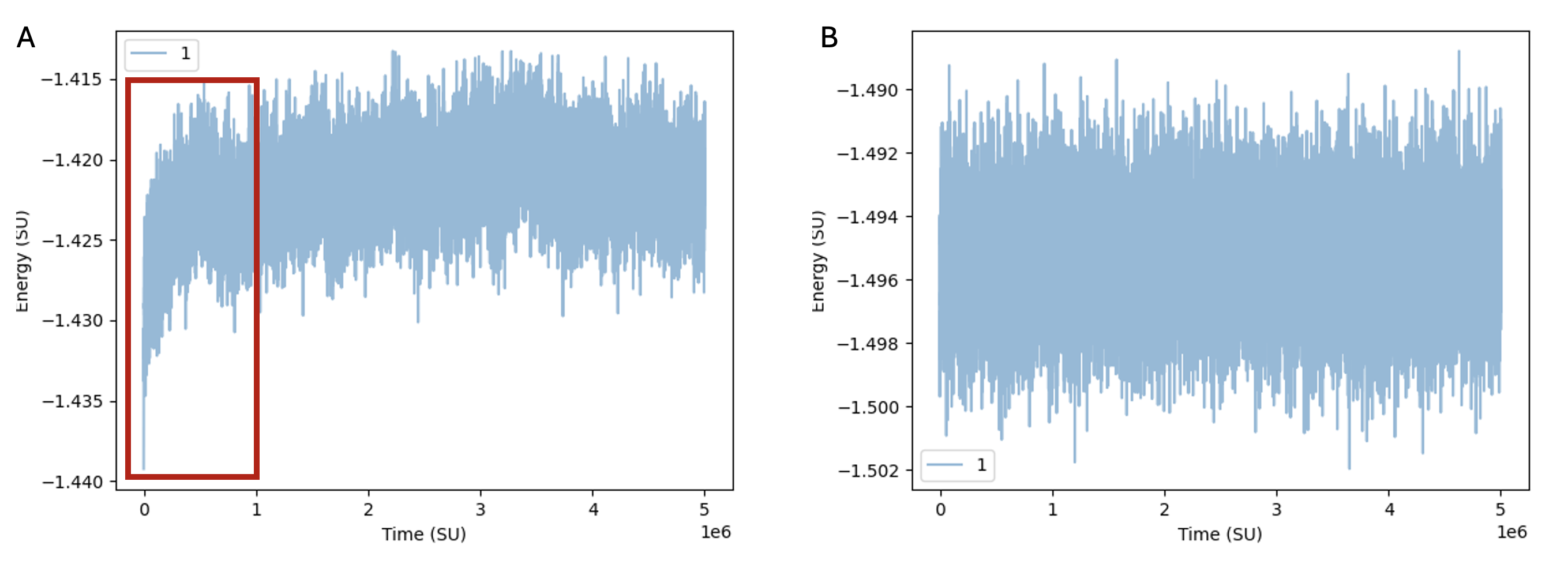}
	\caption{\textbf{Plots of the potential energy of structures throughout the simulation.} (A) Plot of a structure that was not fully equilibrated before running the simulation. Highlighted in red is the beginning of the simulation where the energy still converges towards the end state. (B) Example of the energy plot of a fully equilibrated structure. The energy is only fluctuating around it's mean.}
	\label{fig:energy_plots}
\end{figure}

\subsection{Simulation Production Run}
Aim of the simulation production run is to collect data about the behaviour of the structure. Notice, that before starting the relaxation procedure we applied mutual trap forces between the base pairs in the brick structure to increase stability. However, before starting data collection, these forces must be removed to reflect the structure's natural state. Similarly, the 'Relax settings', which include maximal backbone forces, need to be reset. To reach a stable configuration with those adjusted force parameters, it's crucial to perform an equilibration step before the final production run.
To begin, we delete the forces under the 'Dynamics' > 'Forces' menu. Select all forces by clicking the box next to 'Description' and then press the red 'Delete selected' button. After this step, the list of mutual trap forces should be empty. Next, to run the equilibration, initiate another MD simulation but deselect the 'Use' option under 'Relax settings' in the oxServe settings and adjust the 'dt' values to '0.003'.
We repeat the same for the gear180.
Sufficient equilibration is assessed by monitoring the energy over time. When the potential energy stabilizes (somewhere around -1.4 simulation energy units per particle), the system is likely equilibrated. Additionally, it is crucial to visually check for major structural changes and broken-off fragments. The structures should be intact and stable, moving only slightly around it's equilibrated state.

If this is the case, the structures are ready for the simulation production run. For this we export the structures under 'File' > 'Save' > 'oxDNA'. In the pop-up window we can select the files we want to export and set a filename. Important here is to export the 'Topology' and 'Configuration' files, as those are needed to submit a simulation job on oxDNA.org. After clicking on 'Export' the files should be downloaded as a .top and .dat file. 
The topology file (.top) provides connectivity details, specifying how particles are linked and the sequence assigned to each strand. The header of the topology file indicates the total number of nucleotides and strands. Following, each line corresponds to one nucleotide, being listed from 3' to 5' for each strand. It lists the strand ID, the base identity, and the ID of the nucleotide’s 3' and 5' neighbor. If a nucleotide lacks a 3' or 5' neighbor, the neighbor ID is set to –1. 
The first three lines of the configuration file indicate the timestep, the length of the simulation box sides, and the total, potential, and kinetic energy per particle. Following the header, each nucleotide is described on its own line, which includes the nucleotide’s position, the normal particle orientation vectors, as well as its velocity and angular velocity vectors. Further information about the file format can be found at \href{https://lorenzo-rovigatti.github.io/oxDNA/configurations.html#}{https://lorenzo-rovigatti.github.io/oxDNA/}.

With the saved .top and .dat files we can now submit a simulation at \href{https://oxdna.org}{https://oxdna.org}. 
Users have the option to create an account ('Register') or submit anonymously ('Submit as guest'). Creating an account allows users to opt-in for email notifications and manage all their submissions. Anonymous users, on the other hand, should bookmark the job output link to access their results later.
After registering and verifying their account via email, we can log in and start a simulation by navigating to 'Create A Job'. Here, the topology and configuration files are uploaded, with the option to customize additional parameters as needed. We can name our simulation after 'Job Title' for easier identification later on. While there is an option to relax the structure before the simulation, it is recommended to perform this step in oxView beforehand, where the relaxation process can be monitored and parameters adjusted. Under 'View advanced parameters', set 'dt' to 0.005 and 'average sequence model' to 'no'. For the other parameters the default settings can be kept, and the simulation can be submitted by clicking 'Submit Job'. The duration of the simulation will vary based on the structure's complexity and the number of steps selected, potentially taking several days to complete.  
Results must be downloaded from the server within one week of completion. More details on using the oxdna.org service is provided in Ref.~\cite{poppleton2021oxdna}.

\begin{figure}
	\centering
	\includegraphics[width=\linewidth]{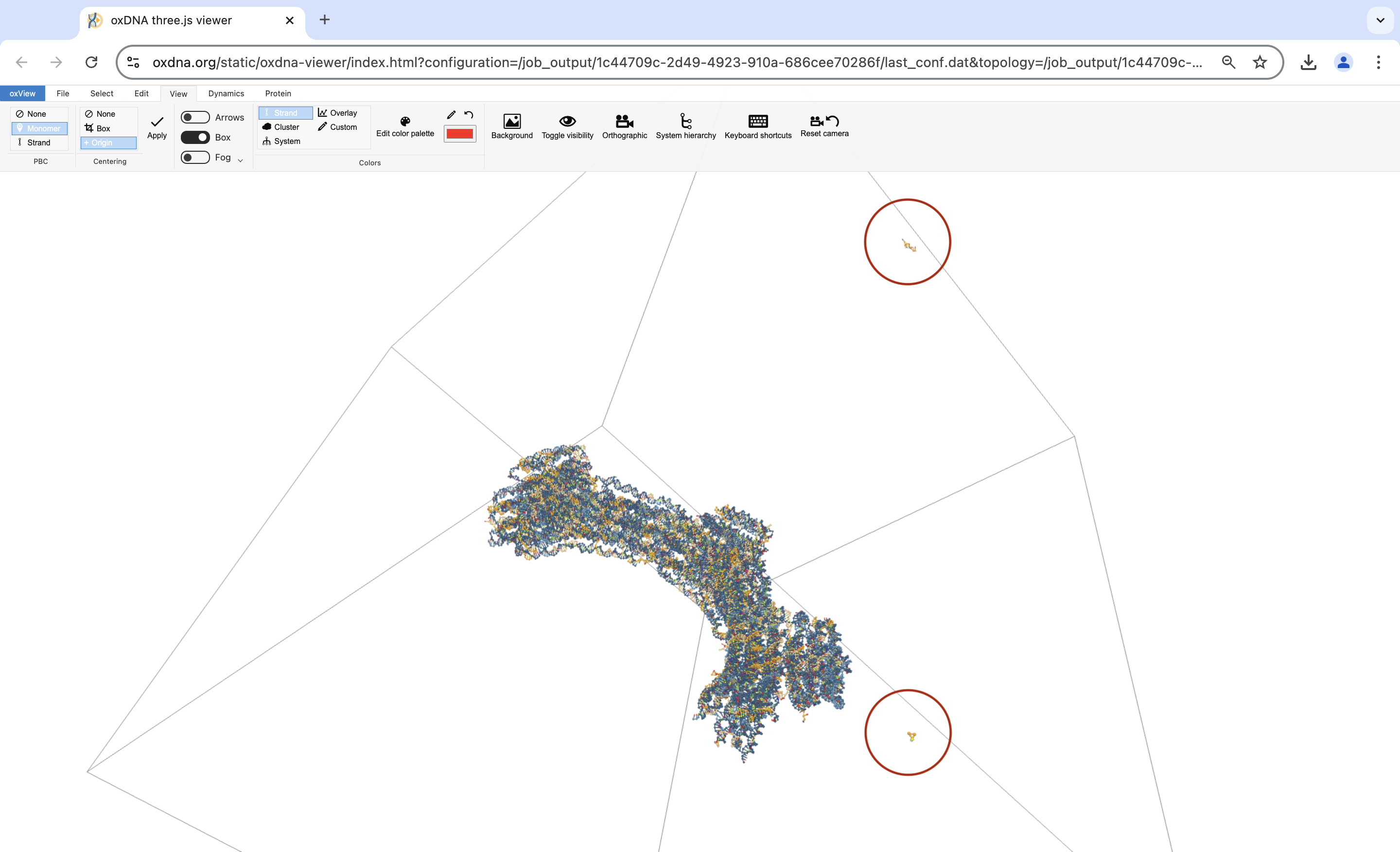}
	\caption{\textbf{Example of a fragmented gear180 structure in the simulation box.} Highlighted in red are two broken off parts.}
	\label{fig:gear180_fragmented_box}
\end{figure}

\begin{figure*}[t]
	\centering
	\includegraphics[width=\linewidth]{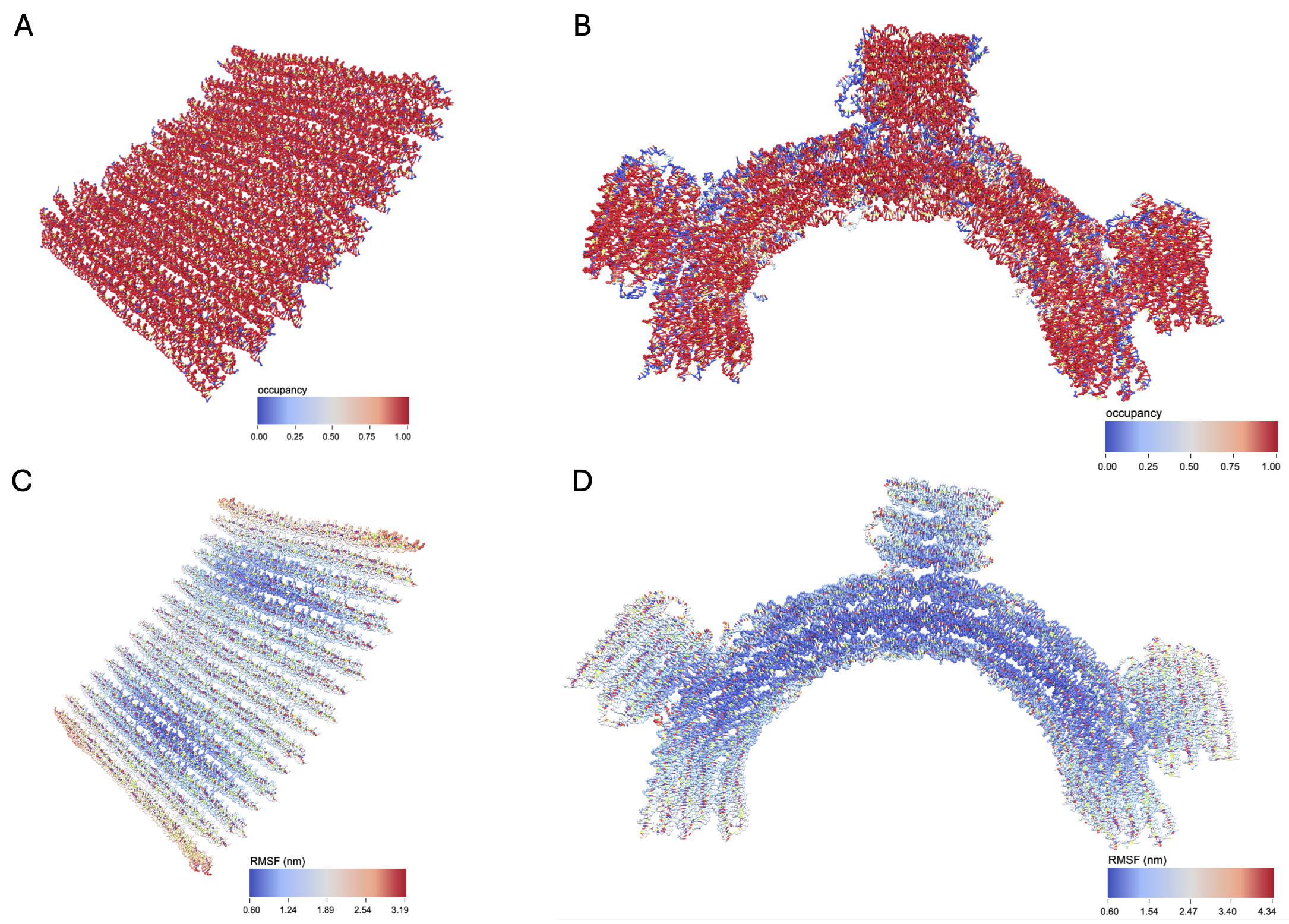}
	\caption{\textbf{RMSF and bond occupancies across the two structures.} (A) Bond occupancy of the brick structure. Dark red areas indicate double-stranded regions, while blue areas indicates single-stranded parts. (B) Bond occupancy of the gear180 structure. In comparison to the tightly bound brick structure, we can observe here multiple single stranded regions. (C) RMSF of the brick structure. Dark blue indicates no fluctuations, while red colored are areas with more movement. (D) RMSF of the gear180 structure. Similar to the brick structure, we observe here that distant regions are more flexible.}
	\label{fig:rmsd_bonds}
\end{figure*}

\subsection{Results Analysis}

As the simulation runs, its progress can be tracked under 'View Jobs', where users can access a list of all jobs. The 'view' option under 'last configuration' opens oxView, allowing a live visualization of the structure to ensure it remains intact.
By clicking on the job name, users are directed to an analysis page. Tools such as 'Mean and RMSF', 'Energy Plotter', and 'Bond Occupancy' offer insights into the structural behavior and fluctuation modes observed during the simulation. The 'Distance' and 'Angles' tools allow for detailed analysis of how different regions of the structure behave in respect to each other. 
While the tools are accessible throughout the simulation process, it is recommended to use them only after it finished.

Upon completion of the simulation it is important to check that the structure stayed intact and was sufficiently equilibrated.
We begin by analyzing the energy plot under 'Energy Plotter' > 'Start Plotting' to confirm that the system's energy has stabilized, showing no significant trends or fluctuations (Figure \ref{fig:energy_plots}). If the energy fluctuates, it indicates that the structure was not fully equilibrated before starting the simulation production run on oxDNA.org. Any data collected before equilibrium is reached should be excluded from further analysis. Additionally, visual inspection of the final configuration in tools like oxView helps to verify structural integrity, ensuring no parts have broken off or are deformed unnaturally. Here it is useful to visualize the box in oxView under 'View' > 'Box' to be able to search the whole area (Figure \ref{fig:gear180_fragmented_box}). If parts were breaking off, the simulation likely failed and the simulation likely failed, and the analysis tools may not provide meaningful results. For further confirmation, tools such as the Root Mean Square Fluctuation (RMSF) and bond occupancy can be used to assess stability throughout the structure.
By clicking on 'View mean+deviations' under the 'Mean and RMSF' section, we can visualize the RMSF across the structure. Dark blue regions represent areas with minimal fluctuation, while light blue highlights nucleotides experiencing greater movement. When examining the structures, we notice light blue or even red coloring at the ends, which is expected as stabilizing cross-overs are less frequent in these distant regions (Figure \ref{fig:rmsd_bonds}C-D). Looking at the bond occupancy under the 'View colormap' option, we can identify distinct structural regions: single-stranded areas appear in blue, strongly bound regions in red, and partially bound regions in white, indicating the dynamic binding and unbinding of staple sequences (Figure \ref{fig:rmsd_bonds}A-B). In the gear180 structure it is clearly visible that the individual helices of the three bricks are connected through single-stranded loops.
Further analyses like calculating distances between nucleotides over time or determining angles can be also performed. 
To visualize the system's dynamics, use the 'Download aligned trajectory' option. Unlike the trajectory file available under 'View Jobs', the aligned trajectory centers the structure within the simulation box and corrects for rotational movements. By dragging and dropping this file along with the corresponding .top file into oxView, the dynamics of the structure can be observed over time under the 'Trajectory' tab. Additionally, videos can be recorded by navigating to 'File' > 'Video'. Alternatively, the structure can be imported into ChimeraX \cite{Meng2023chimeraX} using the tacoxDNA converter, which is available for download at \href{https://github.com/lorenzo-rovigatti/tacoxDNA}{https://github.com/lorenzo-rovigatti/tacoxDNA}.
%All results generated on oxDNA can be downloaded via the provided links.

\section{Conclusion}
In conclusion, we outlined a procedure for simulating DNA origami structures using oxView and oxDNA. However, as highlighted with the gear180 structure, this procedure varies for each specific structure, making it essential to inspect the structure during relaxation and to adjust parameters accordingly. Often overlapping particles and stretched bonds cause numerical instabilities, leading the simulation to crash. With preliminary modifications, applying mutual trap forces during the initial stages and removing them before equilibration, we can achieve effective relaxation of the structure. Key properties, such as potential energy and structural changes are monitored throughout the process until the system is equilibrated. The simulation should then be run for a sufficient duration to ensure proper sampling of the configuration space and to accurately study the structural behavior and dynamics. This duration depends on the dynamics of the system, the complexity of the structure and what accuracy needs to be achieved. Convergence of key properties without long-term trends suggests the simulation has run long enough. Keep in mind that your structures may differ slightly each run due to the inherent stochastic nature of molecular simulations. Components may not align exactly the same in each run, and configurations from relaxation and production runs will vary. However, bulk parameters such as the mean structure and energy should converge to similar values and distributions despite these differences.

Overall, oxView, in combination with oxDNA.org, provides an easy-to use framework to simulate DNA origami structures. Yet, for more complex structures and for in-depth analysis of the structures it is recommended to install the software tool 'oxDNA' and run it locally on a Linux machine, a Linux virtual environment, or a university high performance cluster. For equilibration and proper sampling of more flexible structures or structures that can exist in multiple conformations separated by high energy barriers, advanced sampling methods such as umbrella sampling \cite{sample2023coarse}, metadynamics \cite{kaufhold2022probing} or replica exchange are needed. To extract kinetic information, the forward flux sampling method would need to be applied. More complex analysis of the simulated trajectory that goes beyond the set of tools provided at oxdna.org requires users to write their own python scripts, using the oxDNA analysis tools library. While these advanced tasks are beyond the scope of this introductory paper, they are documented at \href{https://lorenzo-rovigatti.github.io/oxDNA/}{https://lorenzo-rovigatti.github.io/oxDNA/}, along with a detailed installation guide and further information. %, follow the documentation on \href{https://lorenzo-rovigatti.github.io/oxDNA/}{https://lorenzo-rovigatti.github.io/oxDNA/}. 
A more accessible tutorial for installing and launching oxDNA simulations from command line or from Jupyter notebooks either on a local computer or a remote servers is also available in Ref.~\cite{sample2023coarse}.  OxDNA simulations can also be run from Jupyter notebooks using the Google Colab platform. An introductory tutorial using those platforms are provided in the \href{https://colab.research.google.com/drive/1xQutL1crfbPyPBdtfonqNyMQSU1FqP7Z#scrollTo=t9hcM3wzGiSg}{Google Colab} as well as \href{https://tinyurl.com/oxdna}{https://tinyurl.com/oxdna}.

\section*{Acknowledgements}
This result is part of a project that has received funding from the European Research Council (ERC) under the European Union’s Horizon 2020 research and innovation program (Grant Agreement No. 101040035).

%\section*{Conflict of Interest}
%The authors declare no conflict of interest.

\section*{Supporting Information}
 A comprehensive video tutorial accompanying this article can be found at \href{https://youtu.be/5-rgMekX8gE}{https://youtu.be/5-rgMekX8gE}.

\section*{References}
\bibliographystyle{old-mujstyl}  % ama, nar, alpha, plain, chicago, abbrv, siam

\bibliography{refs}

\end{document}